\documentclass[12pt,reqno]{article}
\usepackage{amsmath}
\usepackage{graphicx}
\usepackage{color}
\begin{document}

\title
\centerline{\bf\large Charge density of the electron at higher velocity}

\vspace{12pt}

\centerline{{\bf S. Ghosh}$^{\rm 1}$, {\bf J. K. Sarma } and {\bf N. Pegu}}
\centerline{ Department of Physics, Tezpur University, Tezpur, Assam, 784 028, India} 
\centerline{$^{\rm 1}${\it sovan@tezu.ernet.in}}

\bigskip
\bigskip

\begin{abstract}
\leftskip1.0cm
\rightskip1.0cm
Depending on the recent experiments and some new way of explaining electron we have developed here equations for velocities of different mode with the charge density. This includes as well how the charge-density can be distributed in the equatorial region with different velocities and observer-angles. 
\end{abstract}

\section{Introduction}
\paragraph \ In Standard model of particles electron is taken as a point particle. Quantum mechanically this decision is supported, though till the time a lot of approaches to give a spectroscopic model are going on. With the present situation; i.e. the energy scale available at present days, it is not possible to break the electron, the lightest lepton. But some theoratical endeavours show the possibility of explaining the problems of structural modelling of the electron. One of the leading approaches is relativistically spinning sphere (RSS) model, in which the charge is taken at a point on the equator of the sphere and the charge point is moving around the center of mass of the sphere by zitterbewegung. Though in the RSS model it behaves like a charge point, the latest LEP experiments at CERN suggests that the electron charge is confined within a region of radius $< {10^{-19}}$ m [2]. So the above discussed possibility concludes the fact that the so-called charge-point is moving in a confined region of Compton's wavelength, 
$\lambda_{c}= {\hbar}{/}({mc})\simeq {10^{-13}}$ m. On the other hand the scattering properties of the electron also mandate a vastly smaller radius for electric charge, $R_{e}< {10^{-16}}$ cm [1].
Where as the compton's radius is six order larger than the LEP-given values or the five order of the later one. This essentially reflects the fact that the charge of the electron is concentrated in a place negligibly smaller than the electron volume.

\section{Lienard-Wiechart potential and electron}
\paragraph \ When a charge-point, q is moving on a specified trajectory, the retarded time is determined implicitly by the equation,
\begin{equation}{|{\vec r}-{\vec R'}|}=c(t-T),\end{equation}
where, $\vec R'$ is the retarded position and T is the reatarded time [3]. 
Again $\vec r'= \vec r- \vec R'.$  So the scalar potential is given as ${\phi(\vec r,t)}={{\dfrac{1}{4\pi\epsilon_{0}}}{\int{\frac{\rho(\vec r,T)}{|\vec r-\vec R'|}}dV}},$ and ${\int{\rho(\vec r', T)dV}}={\dfrac{q}{1-\hat r'.({\vec v}{/}{c})}}$.
For convinience we consider the motion of small charged instead of the point-charge. 
We have changed here some notations from fig. 1. to fig. 2. keeping the sense of the figure same.
Then 
\begin{equation}{{\int{\rho(\vec r, T-\dfrac{\vec R}{c})}dV}={\dfrac{q}{1-\dfrac{v}{c}cos\theta}}}.\end{equation} 

Clearly the charge density and hence the scalar potential varies with the angle $\theta,$ which is the angle between the particle velocity and the vector joining the charge retarded position, for v is a constant velocity. For a constant velocity the potential varies with the angle $\theta$ according to Figure 3. At v=c; i.e. the maximum velocity the curve looks like Figure 4. So from Figure 3. and 4., we realize that $\phi$ has a maximum at zero degree and as $theta$ increases, $\phi$ decreases. Hence for any constant velocity v of the charge-part, we have a $\phi_{max}$ and a $\phi_{min}.$ So ${\triangle\phi} = {\phi_{max}-\phi_{min}}.$ From equation (2), ${\int{\rho}dV={\frac{q}{1-\beta cos\theta}}}$ and $\phi=K\int{\rho dV},$ where $K={\frac{1}{4\pi\epsilon_{0}}}=Constant,$ and $\phi=K\int{\rho dV},$ where $K={\frac{1}{4\pi\epsilon_{0}}}= Constant.$ So $\phi= K\int{\rho dV}$ and $\phi=K.{\frac{q}{1-\beta cos\theta}},$ $\phi={\frac{K_{1}}{1-\beta cos\theta}},$ where ${K_{1}}=K.q.$ Again $cos\theta$ ranges from -1 to +1. So ${\phi_{max}}={\frac{K_{1}}{1-\beta}}$ and ${\phi_{min}}={\frac{K_{1}}{1+\beta}}.$ So 

\begin{equation}{{\triangle\phi}={\dfrac{2{K_{1}\beta}}{1-{\beta^{2}}}}}.\end{equation}

\ Or, \begin{displaymath}{{\dfrac{\triangle\phi}{2}}={\dfrac{K_{1}\beta}{1-{\beta^{2}}}}}.\end{displaymath}

\ So ${{{\triangle\phi. \beta^{2}}+{2K_{1}\beta}-{\triangle\phi}}={0}}.$  Hence ${{\dfrac{v}{c}}={\dfrac{\sqrt{K_{1}^{2}+(\triangle\phi)^{2}}-K_{1}}{\triangle\phi}}}.$ Therefore one can write

\begin{equation}{{v_{diff}}={{\dfrac{c}{\triangle\phi}}{[\sqrt{{K_{1}}^{2}+(\triangle\phi)^{2}}-{K_{1}}]}}}.\end{equation}

\ On the other hand, ${\phi_{avg.}={\dfrac{\phi_{max}+\phi_{min}}{2}}={\dfrac{K_{1}}{1-\beta^{2}}}}.$  So 

\begin{equation}{v_{avg.}={c\sqrt{1-{\dfrac{K_{1}}{\phi_{avg.}}}}}}.\end{equation}
So average velocity and the differential velocities are given by equations (4) and (5) which can be concluded from the slope of fig. 3. Using equaion (2), one can draw the graphs of velocity and scalar potential for particular angle $\theta.$

\section{Conclusion}
\paragraph \ From the plot of $\phi$ vs. $\theta$ we can conclude that scalar potential decreases with increasing $\theta.$ Scalar potential is directly related to the charge density. So in other words charge density decreases with increasing $\theta$. This reflects that at smaller angles from the observer the charge density is high. As well as if from different planes this measurement is taken, then at equator the density is high. But, not the entire charge concentration is found at the equator. This is one of the main causes why ${W_E}=0$ [1].

\ Other plots show that scalar potential increases with increasing velocity, after a certain velocity, which is closer to c. So at range-1 in fig.5 potential is independent of v, where as in range-2 it increases in exponential nature. Velocity at boundary of two ranges can be called as critical or boundary velocity.

\ So in the present situation we can conclude that for RSS model the charge location can take place in a region instead of being a point and as we have performed the operation in the equatorial zone, that particular region  matches to our condition already. So this helps us to predict the higher charge-density region in the sphere like shape.

\vspace{100pt}

\section{References}

\ \hspace{20pt}1. M. H. MacGregor, The Enigmatic Electron,  Kluwer Academic 

\ \hspace{15pt} Publishers, Dordrecht (1992).

\ 2. M. Rivas., J. Phys. A: Math. Gen 36 (2003) 4703-4715 , arxiv:physics

\ \hspace{15pt} /0112005 (2003).

\ 3. D. J. Griffiths, Introduction to Electrodynamics, Third edition, 

\ \hspace{15pt} Prentice Hall. Inc., New Delhi (2002).

\ 4. J. M. Aguirregabiria, A. Hernandez, M. Rivas, Am. J. Phys. 60(7), 

\ \hspace{15pt} (1992) 597-599.

\vspace{500pt}

\begin{figure*}
\centerline{
  \mbox{\hspace{65pt} \includegraphics[width=5.50in]{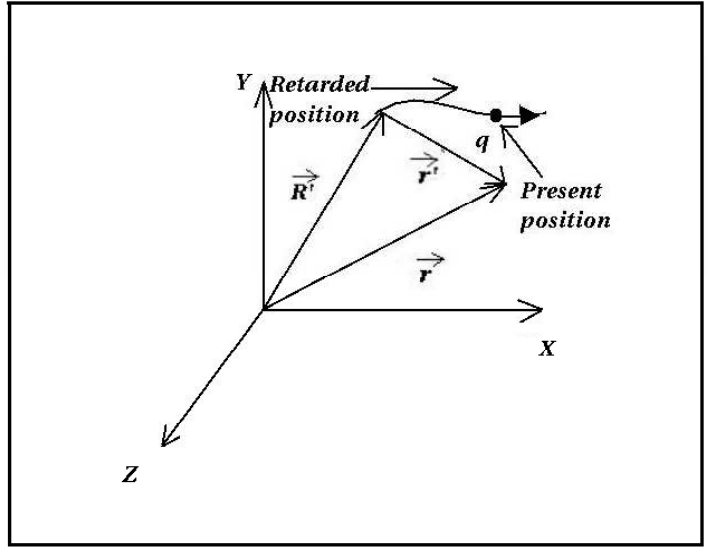}}
  }
    \caption{Particle-travel: retarded position and present position of the moving charge as seen by the external observer.}
\label{overview}
\end{figure*}

\begin{figure*}
\centerline{
  \mbox{\includegraphics[width=5.50in]{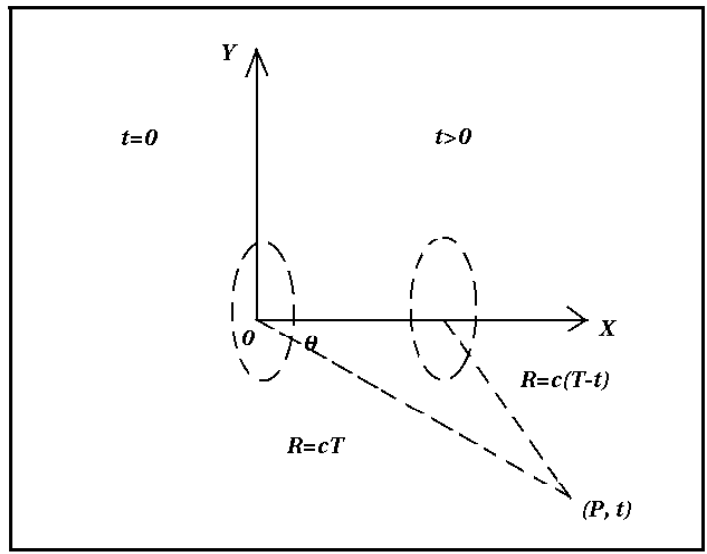}}
  }
    \caption{The intersection of the simultaneous shape of the moving charge and the information collecting sphere.}
\label{overview}
\end{figure*}

\vspace{50pt}

\begin{figure*}
\centerline{
  \mbox{\includegraphics[width=5.50in]{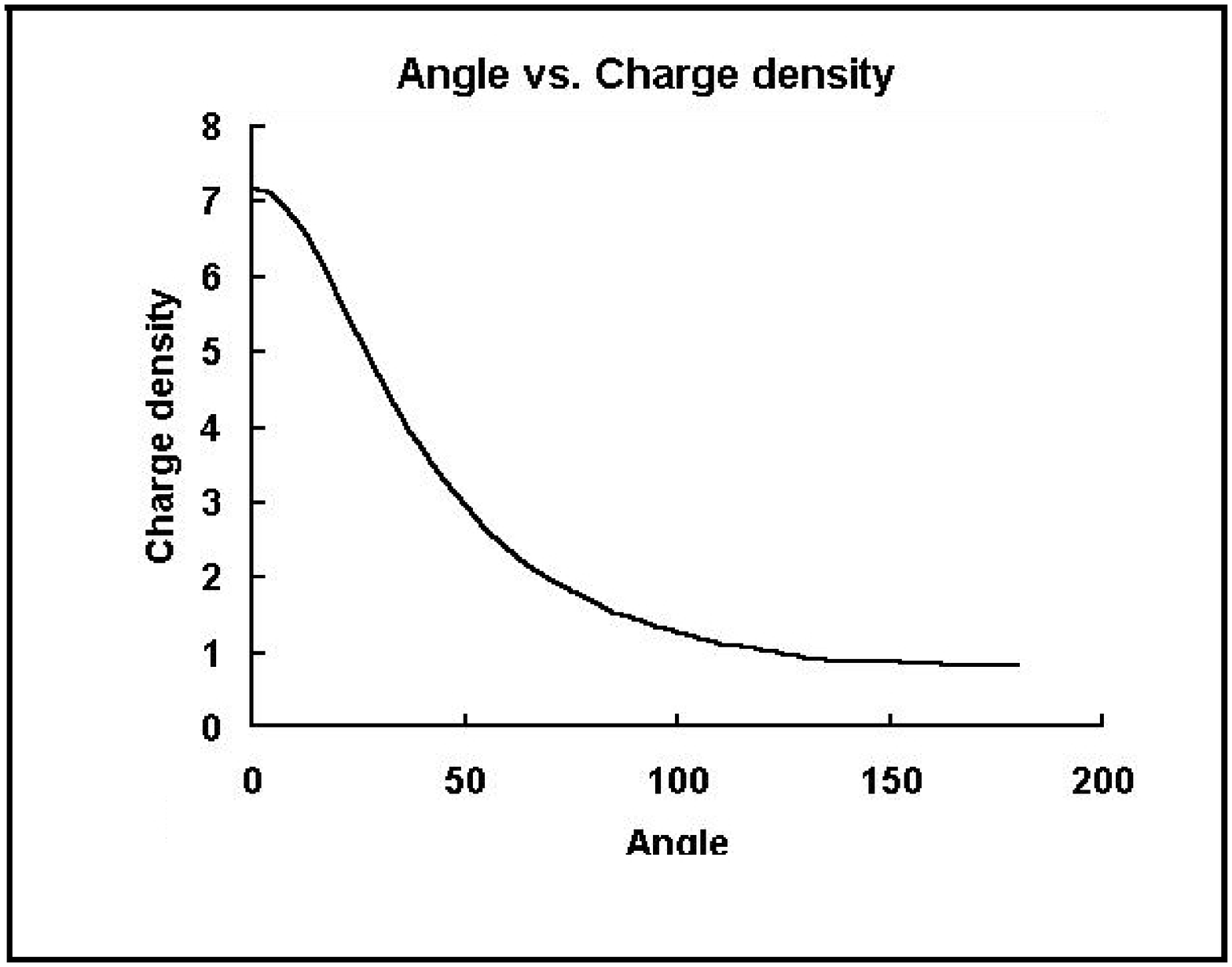}}
  }
    \caption{This is the plot of oberservation angle with charge density; angle at degree and charge density in arbitrary unit.}
\label{overview}
\end{figure*}

\begin{figure*}
\centerline{
  \mbox{\includegraphics[width=5.50in]{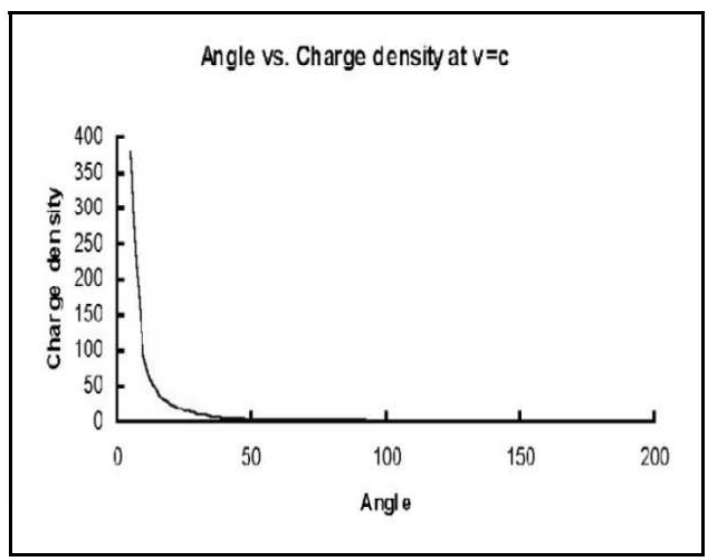}}
  }
    \caption{This is the plot of charge density with angle when v=c; angle at degree and charge density in arbitrary unit.}
\label{overview}
\end{figure*}

\vspace{50pt}

\begin{figure*}
\centerline{
  \mbox{\includegraphics[width=5.00in]{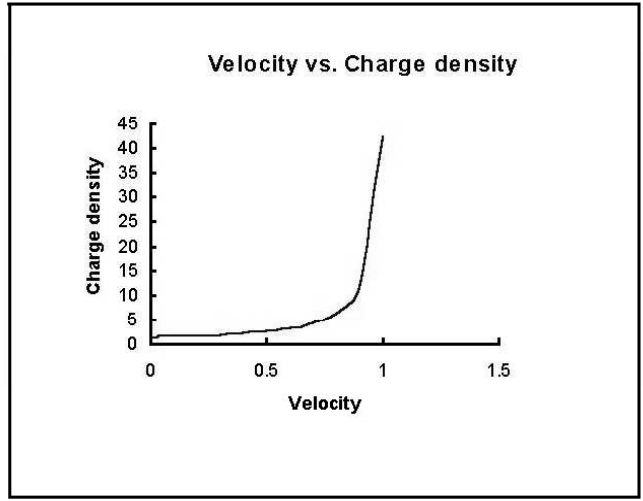}}
  }
    \caption{The plot of velocity with charge density when the observation angle is at first quardrant; velocity in the unit of c and the charge density in arbitrary unit. }
\label{overview}
\end{figure*}

\begin{figure*}
\centerline{
  \mbox{\includegraphics[width=5.00in]{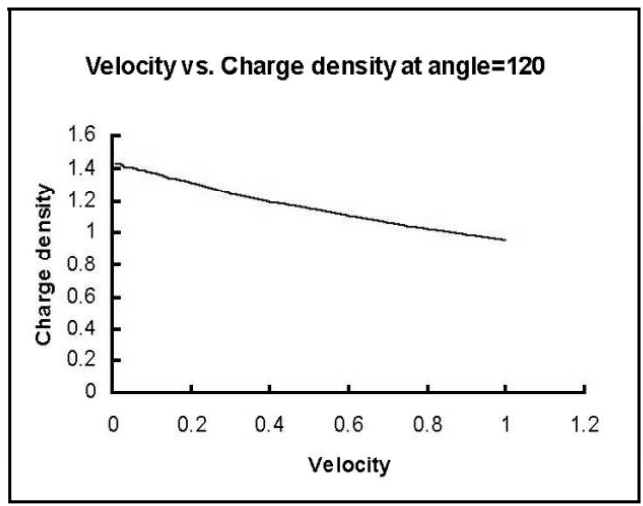}}
  }
    \caption{Here is the plot of velocity with charge density when the observation angle is 120 degree; i.e. at second quardrant; velocity in the unit of c and the charge density in arbitrary unit.}
\label{overview}
\end{figure*}

\end{document}